\documentclass[conference]{IEEEtran}
\ifCLASSINFOpdf
\else
\fi
\usepackage{cite}
\usepackage{amsmath,amssymb,amsfonts}
\usepackage{algorithmic}
\usepackage{graphicx}
\usepackage{textcomp}
\usepackage{comment}
\usepackage[nolist]{acronym}
\usepackage[nolist]{acronym}
\usepackage[normalem]{ulem}
\usepackage{comment}
\usepackage{color,soul}
\usepackage{multirow, makecell}
\usepackage{cellspace, booktabs}
\usepackage{multicol,xcolor,colortbl}
\usepackage{tikz,pgfplots}
\pgfplotsset{width=10cm,compat=1.9}
\usepackage{adjustbox}
\usepackage{scalefnt}
\usepackage{balance}
\ifCLASSOPTIONcompsoc
    \usepackage[caption=false, font=normalsize, labelfont=sf, textfont=sf]{subfig}
\else
\usepackage[caption=false, font=footnotesize]{subfig}
\fi
\usepackage{placeins}
\usepackage{orcidlink}
\usepackage{dblfloatfix}

\newcommand{\red}[1] {\textcolor[rgb]{1.0,0.0,0.0}{{#1}}}

\newcommand{\pkeep}{p_\text{k}}

\usepackage{longtable}
\usepackage{threeparttable}
\newcolumntype{L}[1]{>{\raggedright\let\newline\\\arraybackslash\hspace{0pt}}m{#1}}
\newcolumntype{C}[1]{>{\centering\let\newline\\\arraybackslash\hspace{0pt}}m{#1}}
\newcolumntype{R}[1]{>{\raggedleft\let\newline\\\arraybackslash\hspace{0pt}}m{#1}}
\newcolumntype{M}[1]{>{\centering\arraybackslash}m{#1}}

\begin{document}
%
\title{	Can NR-V2X Sidelink support A2A links?}

\author{\IEEEauthorblockN{
Vittorio Todisco\IEEEauthorrefmark{1}\IEEEauthorrefmark{2}\orcidlink{0000-0002-6737-6710},
Alessandro Bazzi\IEEEauthorrefmark{1}\IEEEauthorrefmark{2}\orcidlink{0000-0003-3500-1997}
}
\medskip
\IEEEauthorblockA{\IEEEauthorrefmark{1}DEI, Universit\`a di Bologna, 40136 Bologna, Italy}
\IEEEauthorblockA{\IEEEauthorrefmark{2}National Laboratory of Wireless Communications (WiLab), CNIT, 40136 Bologna, Italy}
\thanks{Corresponding author: V. Todisco (email: vittorio.todisco@unibo.it).}}


\markboth{Journal of \LaTeX\ Class Files,~Vol.~14, No.~8, August~2015}%
{Shell \MakeLowercase{\textit{et al.}}: Bare Demo of IEEEtran.cls for IEEE Transactions on Magnetics Journals}
%




\begin{acronym} 
\acro{3GPP}{Third Generation Partnership Project}
\acro{4G}{fourth-generation}
\acro{5G}{fifth-generation}
\acro{6G}{sixth-generation}
\acro{5GAA}{5G Automotive Association}
\acro{A2A}{Air-to-Air}
\acro{AF}{amplify and forward}
\acro{AGC}{automatic gain control}
\acro{AoI}{age of information}
\acro{AR}{augmented reality}
\acro{AS}{application server}
\acro{AWGN}{additive white Gaussian noise}
\acro{B-CSA}{broadcast coded-slotted ALOHA}
\acro{BSM}{basic safety message}
\acro{BS}{base station}
\acro{C-NOMA}{cooperative NOMA}
\acro{C-V2X}{cellular-V2X}
\acro{CAM}{cooperative awareness message}
\acro{CAV}{connected and autonomous vehicle}
\acro{CBR}{channel busy ratio}
\acro{CCDF}{complementary cumulative distribution function}
\acro{CDF}{cumulative distribution function}
\acro{CD}{collision detection}
\acro{CoMP}{coordinated multi-point}
\acro{CP-OFDM}{cyclic prefix orthogonal frequency-division multiplexing}
\acro{CPM}{collective perception message}
\acro{CRC}{cyclic redundancy check}
\acro{CRDSA}{contention resolution diversity slotted ALOHA}
\acro{CSA}{coded-slotted ALOHA}
\acro{CSAC}{Chip scale atomic clock}
\acro{CSI}{channel state information}
\acro{CSIT}{channel state information at the transmitter}
\acro{CSIR}{channel state information at the receiver}
\acro{CSMA/CA}{carrier sense multiple access with collision avoidance}
\acro{D2D}{device-to-device}
\acro{DCI}{downlink control information}
\acro{DENM}{decentralized environmental notification message}
\acro{DIT}{Double Initial Transmission}
\acro{DMRS}{demodulation reference signal}
\acro{DS}{dynamic scheduling}
\acro{ECP}{extended cyclic prefix}
\acro{EED}{end-to-end delay}
\acro{eNodeB}{evolved NodeB}
\acro{ETSI}{European Telecommunications Standards Institute}
\acro{FD}{in-band full-duplex}
\acro{GNSS}{global navigation satellite system}
\acro{HD}{half-duplex}
\acro{IBE}{in-band emission}
\acro{ICI}{Inter-carrier interference}
\acro{IDMA}{interleave-division multiple access}
\acro{IM}{index modulation}
\acro{ISI}{inter-symbol interference}
\acro{JD}{joint decoding}
\acro{KPI}{key performance indicator}  
\acro{LDPC}{low-density parity-check}
\acro{LOS}{line-of-sight}
\acro{LoA}{Level of Automation}
\acro{LTE}{long-term evolution}  
\acro{LTE-V2X}{long-term evolution-V2X}
\acro{MAC}{medium access control}
\acro{MCM}{maneuver coordination message}
\acro{MCS}{modulation and coding scheme}
\acro{MIMO}{multiple input multiple output}
\acro{mmWave}{millimiter wave}
\acro{MRC}{maximum ratio combining}
\acro{MUD}{multi-user detection}
\acro{NLOS}{non-line-of-sight}
\acro{NOMA}{non-orthogonal multiple access}
\acro{NTN}{Non-Terrestrial Networks}
\acro{NR}{New Radio}
\acro{NR-V2X}{New Radio vehicle to everything} 
\acro{OBU}{on-board unit}
\acro{OCDM}{orthogonal chirp-division multiplexing}
\acro{OFDM}{orthogonal frequency-division multiplexing}
\acro{OFDMA}{orthogonal frequency-division multiple access}
\acro{OMA}{orthogonal multiple access}
\acro{OTFS}{orthogonal time frequency space}
\acro{PAPR}{peak to average power ratio}
\acro{PCM}{platoon control message}
\acro{PDB}{packet delay budget}
\acro{PER}{packet error rate}
\acro{PIR}{packet inter-reception}
\acro{PSCCH}{Physical Sidelink Control Channel}
\acro{PSFCH}{Physical Sidelink Feedback channel}
\acro{PSSCH}{Physical Sidelink Shared channel}
\acro{PHY}{physical}
\acro{PRB}{physical resource block}
\acro{PRR}{Packet Reception Ratio}
\acro{QAM}{quadrature amplitude modulation}
\acro{QoS}{quality of service}
\acro{QPSK}{quadrature phase shift keying}
\acro{RC}{reselection counter}
\acro{RF}{radio frequency}
\acro{RRI}{resource reservation interval}
\acro{RSRP}{Reference Signal Received Power}
\acro{RSU}{road side unit} 
\acro{SB-SPS}{sensing-based semi-persistent scheduling}
\acro{SB-DS}{sensing-based dynamic scheduling}
\acro{SCH}{service channel}
\acro{SCI}{sidelink control information}
\acro{SCI-RR}{resource reselection based on SCI decoding and re-evaluation}
\acro{SCMA}{sparse-code multiple access}
\acro{SCS}{subcarrier spacing}
\acro{SI}{self-interference}
\acro{SIC}{self-interference cancellation}
\acro{SINR}{signal-to-interference-plus-noise ratio}
\acro{SNR}{signal-to-noise ratio}
\acro{SPS}{semi-persistent scheduling} 
\acro{SR}{scheduling request}
\acro{TB}{transport block}
\acro{TDMA}{time-division multiple access}
\acro{TTI}{transmission time interval}
\acro{UAV}{unmanned aerial vehicle}
\acro{UE}{user equipment}
\acro{URLLC}{ultra-reliable and ultra-low latency communications}
\acro{V2N}{vehicle-to-network}
\acro{V2I}{vehicle-to-infrastructure} 
\acro{V2P}{vehicle-to-pedestrian}
\acro{V2V}{vehicle-to-vehicle} 
\acro{V2X}{vehicle-to-everything} 
\acro{VRU}{vulnerable road user}
\acro{VUE}{vehicular user equipment}
\acro{WBSP}{Wireless Blind Spot Probability}
\end{acronym}

\IEEEtitleabstractindextext{%
\begin{abstract}

In the context of 5G, 3GPP introduced  \ac{NR-V2X} for direct vehicle-to-vehicle communication. However, starting from Release 18 the focus of the standard has been expanded from vehicles to any device and use case that can benefit from direct communication. In 3GPP terminology, the standard is now referred to simply as Sidelink communication. This standard allows direct communication between devices based on synchronous resource scheduling. Users can rely on controlled scheduling when in network coverage or, in the absence of coverage, autonomously select resources for transmission via a distributed resource allocation mechanism. Focusing on the autonomous resource allocation, this paper investigates the possibility of applying  Release 18 Sidelink communication to \ac{A2A} links between airborne entities. The paper outlines the main challenges and required modifications to adapt the current standard for longer links in the order of kilometres. The analysis identifies the propagation delay as a critical limitation. Communications at distances over 42.4~km require a restriction of the user's transmitting opportunities. However, sidelink communication remains feasible for distances below this threshold without modifications to the standard.

\end{abstract}

}

\maketitle

\IEEEdisplaynontitleabstractindextext

%
\IEEEpeerreviewmaketitle

\acresetall

\section{Introduction}
Telecommunication has long been rooted in two-dimensional frameworks, predominantly relying on ground-based infrastructure like cell towers and fibre-optic cables. However, the growing interest for \ac{UAV} based services and \ac{NTN} may lead to a new era where telecommunication networks are evolving into three-dimensional structures. This transformation is driven by the integration of nodes such as satellites, drones, and high-altitude platforms into the communication grid.

The inclusion of these advanced nodes in telecommunication networks enhances global coverage and connectivity. UAVs offer agile and flexible communication nodes capable of rapid deployment and dynamic network configurations \cite{Mignardi_Drones}. High-altitude platforms, hovering in the stratosphere, provide persistent and wide-area coverage \cite{9773096}. Satellites in NTN extend coverage to remote and underserved areas where traditional ground-based infrastructure is impractical or too costly. This expanded coverage not only improves accessibility but also supports applications in navigation, weather monitoring, and disaster management \cite{Bajracharya_Drones}.

This shift to 3D networks opens up a realm of possibilities and new use cases, including enhanced Internet of Things (IoT) connectivity, improved latency for real-time applications, and the seamless integration of terrestrial and aerial communication systems \cite{Conserva_Drones,Mignardi_Drones2,Spampinato_Drones,Mishra6G}. This multidimensional telecommunication landscape promises to revolutionize how we connect, communicate, and interact on a global scale.

However, as a result of this integration, there will be an increasing number of objects occupying the airspace, leading to a more crowded and dynamic environment. This proliferation of airborne entities necessitates reliable and efficient communication systems to ensure safety, coordination, and optimal use of airspace. As the number of communicating objects in the airspace grows, it is necessary to address how these entities access and share the wireless channel \cite{Bajracharya_Drones}.

In some use cases, these entities need to communicate directly in a distributed manner, leading to the concept of \ac{A2A} communication. A2A communication enables the direct exchange of information between airborne entities without relying on intermediary ground stations or satellite links. This direct communication is essential for several key applications, such as collision avoidance, formation flying, real-time information sharing, and coordinated maneuvers.
%
%
\begin{figure}[t]
\centering
\includegraphics[width=\columnwidth]{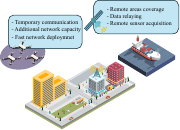}
\caption{Example of the hybrid 3D network and key use cases.}
\label{fig:persColl}
\end{figure}

Moreover, direct communication is beneficial as it can reduce the latency and it can also improve the reliability of the communication by reducing the dependency on the availability of the infrastructure \cite{Bajracharya_Military}.

Recognizing the importance of such aspects, the \ac{3GPP} expanded support for direct communication in Release 16 as part of 5G \ac{NR-V2X}. This standard enables direct \ac{D2D} communication, allowing vehicles to communicate directly, either through a distributed procedure or with support from the infrastructure. 
In Release 18, \ac{3GPP} broadened the standardization effort beyond the vehicular sector to include a wide range of devices and use cases, reflecting the broad applicability of direct communication. The standard is now referred to simply as Sidelink communication.

This work examines the applicability and limitations of the NR Release 18 Sidelink autonomous mode,for A2A links between aerial vehicles over long distances, ranging from a few kilometres to 150 km.


\smallskip

The paper is organized as follows.
Section~\ref{sec:literature} summarises the relevant literature regarding Sidelink and airborne entities.
Section~\ref{sec:DirectCom} overviews the direct communication paradigm and its benefits. Section~\ref{sec:Sidelink} reviews the Sidelink standard and provides details on the resource allocation and the synchronization procedure.
The applicability of Sidelink to \ac{A2A} is discussed in Section~\ref{sec:analysis}. Section~\ref{sec:conc} concludes the paper. 

\section{Related work on Sidelink for A2A links}\label{sec:literature}


To the best of our knowledge, very few studies have investigated the applicability of the Sidelink standard to airborne devices like drones, planes, and satellites. Most studies focus on the particular use case of cellular-connected \acp{UAV}, where drones are served from the cellular network as surveyed in \cite{mishra2}. \cite{spectrumSharing} showed difficulties in providing connectivity for drones as the cellular antennas are generally designed to orient the main lobe toward the ground, resulting in UAVs being served through the antenna \ac{BS} side lobes with reduced power.
Only a few works, such as \cite{mishra1,corridors,MishraC-U2X}, focus on sidelink applications. In \cite{mishra1}, the idea of leveraging the sidelink to enable connectivity among a group of \acp{UAV} is introduced and three different possible sidelink-assisted multi-hop routing protocols to deliver information between the users are discussed. 
\cite{corridors}, explores the concept of drone corridors which are designated airspaces where UAVs can operate in an organized and structured manner, similar to air traffic lanes. The authors argue that drone corridors coupled with millimetre wave and directive antennas reduce reciprocal interference over traditional isotropic antennas. 
\cite{MishraC-U2X},
considers a swarm of UAVs scheduling data transmission from a set of target points of interest towards the cellular \ac{BS}. The authors introduce a Sidelink cooperative multi-hop
communication model which can expand the UAV to infrastructure communication range.
However, none of the aforementioned works have comprehensively analyzed the various aspects of sidelink technology and their interplay in the design of communication links for airborne entities.



\section{Direct communication paradigm}\label{sec:DirectCom}



Traditional telecommunication networks have primarily relied on uplink and downlink communication schemes, where devices communicate with each other indirectly through a central base station or network infrastructure. This model has been effective for many applications, but it also introduces several limitations, such as increased latency and dependency on network availability and reliability.
The use of direct communication, also known as \ac{D2D} communication, represents a paradigm shift. In D2D communication, devices can communicate directly with each other without the need for an intermediary.
Since the wireless channel is a shared medium, different protocols have been developed to manage scenarios where multiple users may compete to access the channel. Some of the key standards include Wi-Fi Direct, ZigBee and the 3GPP Sidelink, the focus of this work.
%


\begin{figure}[t]
\centering
\includegraphics[width=0.90\columnwidth]{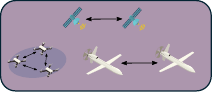}
\caption{Example of applications of direct communications between air entities.}
\label{fig:useCasesD2D}
\end{figure}
%

Fig. \ref{fig:useCasesD2D} portrays possible applications where airborne entities could benefit from direct communication, while the following discusses some of the key benefits. 
%
\smallskip
\\\textbf{Infrastructure Availability Independence:} Devices with direct communication capabilities can establish communication links without relying on existing network infrastructure. This independence from fixed infrastructure allows for greater flexibility and rapid deployment, making it possible to operate in remote areas. Moreover, 
direct communication ensures that devices can continue to operate even if the primary network becomes unavailable due to natural disasters, technical failures, or other disruptions.
\\\textbf{Improved Latency:} Direct communication significantly reduces latency by allowing data to be exchanged directly between devices without being relayed from the network. This low-latency communication is essential for real-time applications such as coordinated flight, formation flying, and dynamic mission adjustments.
%
%
\\\textbf{Efficient Use of Resources:} D2D communication offloads traffic from the central network, thereby increasing overall network efficiency and reducing congestion. This is beneficial in densely populated areas or during large events where network demand spikes.
\\\textbf{Localized Communication:} In scenarios where data is relevant only within a specific locality, such as environmental monitoring, search and rescue operations, or localized surveillance, D2D communication ensures that information is disseminated quickly and efficiently within the local context.
%
Moreover, with fewer intermediaries, there are fewer points of potential interception or attack, enhancing the security and privacy of the communication.

\smallskip

Given these advantages, direct communication via D2D protocols may be a relevant component of modern airborne communication networks, enabling a wide range of applications from disaster response and environmental monitoring to 
drone swarm operations and commercial services.

\section{3GPP Sidelink}\label{sec:Sidelink}

Hereafter, the main aspects of the 3GPP Sidelink standard are recalled. For a more detailed description of the 3GPP specifications, the reader is referred to \cite{todisco2021performance,GarMolBobGozColSahKou:21,from4Gto5G}.

\subsection{Resource Grid}\label{sec:resGrid}

In Sidelink communication, all \acp{UE} are assumed to be synchronized 
and share the available bandwidth, organized as a grid of 
time and frequency resources. Time resources are called slots and their duration is fixed to a fraction of a millisecond, depending on the numerology $\mu\in[0,1,2,3]$ (e.g., $2^{-\mu}$~ms when the \ac{SCS} equals $15\cdot2^\mu$ kHz). Frequency resources, on the other hand, are divided into subchannels, which are groups of contiguous subcarriers. Each packet is allocated and transmitted in a single 
slot and a number of subchannels, depending on its size and the adopted \ac{MCS}.

\subsection{Sidelink Resource Allocation}

Different protocols have been designed over the years to optimize the access to a shared channel from multiple distributed nodes. Traditional procedures include slotted Aloha and listen-before-talk methods like \ac{CSMA/CA}. 
The Sidelink protocol differs from these traditional standards by considering a more structured resource allocation mechanisms derived from the established practices of cellular networks. Sidelink devices can inherently leverage the use of a central scheduler or revert to a distributed mechanism when necessary, providing flexibility and robustness in various communication scenarios.
Two resource allocation modes are defined:
\\\textbf{Mode~1:} In mode~1, the network (typically a base station) schedules the resources for sidelink communication. This mode relies on centralized control, where the base station allocates specific time and frequency resources to each device for direct communication.
\\\textbf{Mode~2:} In mode~2, devices autonomously select and manage their communication resources without relying on a central network controller. Each device monitors the channel to detect future time/frequency reservations from other users.

\subsection{Autonomous Mode}
When the autonomous mode 
is used, the resource allocation triggered by the generation of a new packet is performed directly by each UE. Each UE identifies the set of available resources $S_a$ through two steps: \textit{(i)} channel sensing; \textit{(ii)}
resource exclusion. Then, the obtained $S_a$ is passed to the higher layers, which proceed to select a resource from within the set.
In the first step, the \ac{UE} considers a sensing window covering the last 1.1~or 0.1~seconds; it reads the \ac{SCI} received with the decoded packets in that interval. 
Specifically, the \ac{SCI} may reserve resources in the future. In the second step, the UE discards resources from inside a future interval called selection window. The selection window ends after a given time relative to the packet generation, set according to the delay budget. 

SCI-reserved resources are discarded if the associated received power exceeds a given threshold. If fewer resources than a predetermined threshold are available, the procedure is repeated with an increased received power threshold. This adjustment may result in the exclusion of reservations from more distant UEs. 

\subsection{Resource Scheduling}
After obtaining the set of available resources, $S_a$, the higher layers select the resource for transmission. Access to the channel is provided according to two possible resource scheduling mechanisms. 
Two approaches are therefore available for handling the traffic generated by the UE's applications: the \ac{SPS} and the \ac{DS}. 
In the \ac{SPS} the UE periodically reserves a time-frequency resource for transmission with a time interval equal to the \ac{RRI}. The SPS accesses the resource several times according to the \ac{RC} whose value is randomly initialized. When the RC reaches expires, a reselection is performed with a probability $p_k$, and the \ac{RC} is reinitialized. As discussed in the literature, the \ac{SPS} is more suitable for periodic traffic. Differently, with the \ac{DS}, the UE selects new resources for every new packet, which is preferable to serve aperiodic traffic \cite{lusvarghi2023comparative}.

\subsection{Synchronization}

For the correct operation of the Sidelink system, synchronization between all the communicating nodes is required. Sidelink synchronization refers to the process of obtaining precise timing references for the frame and symbol boundaries. This synchronization ensures that all devices have a common temporal reference, enabling the possibility of scheduling the access to the communication channel.

In NR Sidelink, synchronization can be derived from four primary sources: a GNSS source, a base station (gNB), another UE, or the UE's own internal clock.

GNSS, such as GPS, provides highly accurate timing information by solving the trilateration problem, which requires signals from at least four satellites. When a UE is synchronized with GNSS, it obtains both precise timing and positional information. 
The base station, or gNB, can also serve as a reliable source of synchronization. The UE decodes specific synchronization signals, known as the Synchronization Signal Block (SSB), transmitted by the gNB. This method is particularly reliable in areas with good network coverage.
In the absence of GNSS or gNB synchronization, a UE can derive timing information from another UE that is already synchronized. This UE, known as a SyncRef UE, becomes the reference for other UEs within its communication range. SyncRef UEs are categorized based on their synchronization source, which can be directly from GNSS/gNB or indirectly through other UEs.
Finally, when no external synchronization reference is available, a UE resorts to its own internal clock. Although this is the least preferred method, it ensures that the UE can still operate in isolated scenarios, albeit with potentially reduced synchronization accuracy.

\section{Sidelink Analysis}\label{sec:analysis}
This section examines key factors influencing the design of sidelink communication between airborne devices. It highlights critical aspects and potential countermeasures.

\subsection{Communication range}

The link budget is the first step in designing a communication link, it is essential for determining whether a communication link can be successfully established and maintained under specific conditions. The first step is the determination of a suitable channel model. For what regards the Sidelink for vehicular and urban cellular scenarios different models have been designed. Such models may take into consideration the effects of multipath propagation, shadowing from buildings and vehicles, which characterize the urban environment.
However, for \ac{A2A} communication, especially at higher altitudes, the channel characteristics differ significantly from terrestrial environments. Aerial devices move in the 3D space with trajectories and relative speeds that can be very different from the ones of vehicles. Moreover, effects such as rain attenuation and atmospheric attenuation may become significant, especially for long-distance links.  However, the absence of large obstacles and the \ac{LOS} nature of A2A links lead to a simpler channel model, primarily dominated by free-space path loss.
Nevertheless, the free space path loss model is valid only when there is an unobstructed \ac{LOS} path between the transmitter and the receiver and no objects in the first Fresnel zone. This condition may not be true for a drone swarm, where a group of drones fly together in proximity. \cite{channelMod1} discusses such aspects considering neighbouring drones as scatterers in the proposed channel model.
For more details on channel modelling for avionic links, refer to the survey \cite{channelModSurvey}.
As this article focuses on long links a free-space path loss model will be considered, modelled according to Eq. \ref{eq:fspl}.

\begin{equation}
PL_{\text{dB}}(d) = 32.4 + 20 \log_{10}(d) + 20 \log_{10}(f_0)
\label{eq:fspl} \;.
\end{equation}

\begin{table}[b]
\caption{Parameters for maximum link length evaluation.}
\vspace{-2mm}
\label{Tab:LinkBudget}
\footnotesize
\centering
\begin{tabular}{ll}
\toprule
\midrule
Carrier Frequency & $6$~GHz \\
Bandwidth & $100$~MHz \\
SCS & $60$~kHz \\
TX Power & $40$~dBm \\
TX Gain & $12$~dBi \\
RX Gain & $12$~dBi \\
RX Noise Figure, $F_d$ & $8$~dB \\
\toprule
\midrule
\end{tabular}
\end{table}

To determine the coverage range for sidelink transmissions in \ac{A2A} links, several assumptions about the communication link chain must be made. The characteristics of the link parameters are summarized in Table \ref{Tab:LinkBudget}. The carrier frequency is set to $6$~GHz, and the maximum bandwidth sidelink bandwidth of $100$~MHz is considered. The \ac{SCS} is set to $60$~kHz, the maximum value allowed by the standard for applications below $7$~GHz. This choice will be discussed in more detail in a subsequent section.

The transmit power is set to $40$~dBm, with both the transmitter and receiver antenna gain set to $12$~dBi. The selection of a relatively directive antenna is justified in drone applications, where multiple antennas are typically colocated at different positions on the aircraft. This configuration helps to minimize the likelihood of blockage and \ac{LOS} obstruction caused by parts of the airplane itself.

With the parameters considered and assuming free-space propagation in \ac{LOS} conditions, the received power at a given distance can be evaluated according to the Friis transmission equation. The minimum received power to have a successful transmission $P_\text{rmin}$,is evaluated as the minimum power above the \ac{SNR} threshold and the combined noise level, taking into account the Boltzmann constant $k$, the thermal noise $T_\text{0}$, the bandwidth $B_\text{W}$, and the receiver noise figure $F_\text{R}$.

\begin{equation}
P_\text{rmin} = SNR_\text{min} + 10\log_{10}(k \cdot T_\text{0} \cdot B_\text{W}) + F_\text{R}
\end{equation}


\definecolor{mycolor1}{rgb}{0.00000,0.44700,0.74100}%
\definecolor{mycolor2}{rgb}{0.85000,0.32500,0.09800}%
\definecolor{mycolor3}{rgb}{0.92900,0.69400,0.12500}%
\definecolor{mycolor4}{rgb}{0.49400,0.18400,0.55600}%
\definecolor{mycolor5}{rgb}{0.46600,0.67400,0.18800}%
\definecolor{mycolor6}{rgb}{0.30100,0.74500,0.93300}%
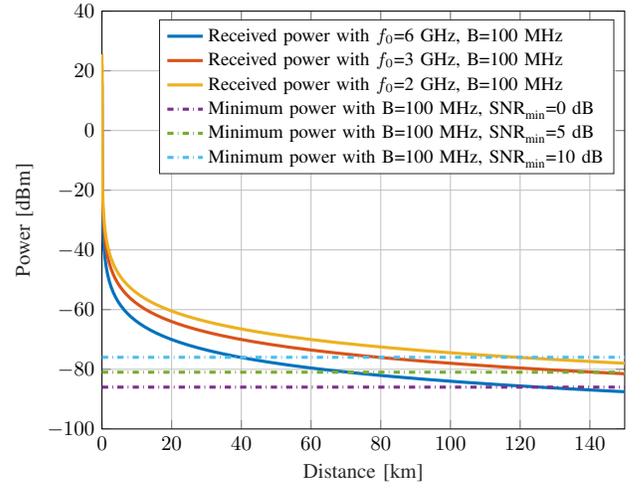
\begin{figure}[h!]
        \centering
\begin{adjustbox}{width=0.925\columnwidth,keepaspectratio}

\begin{tikzpicture}

\begin{axis}[%
width=\columnwidth,
height=0.8\columnwidth,
scale only axis,
xmin=0,
xmax=150,
xlabel style={font=\color{white!15!black}},
xlabel={Distance [km]},
ymin=-100,
ymax=40,
ylabel style={font=\color{white!15!black}},
ylabel={Power [dBm]},
axis background/.style={fill=white},
xmajorgrids,
ymajorgrids,
legend style={legend cell align=left, align=left, draw=white!15!black, font=\small}
]
\addplot [color=mycolor1, line width=1.5pt]
  table[row sep=crcr]{%
0.00100000000000477	15.9952028062785\\
0.0600000000000023	-19.5678222013944\\
0.179000000000002	-29.0618578133194\\
0.355999999999995	-35.0337971531791\\
0.588999999999999	-39.4071030894636\\
0.878999999999991	-42.884574695197\\
1.226	-45.7746065973695\\
1.63	-48.2485492818007\\
2.09299999999999	-50.4201817604953\\
2.61600000000001	-52.3575519867661\\
3.202	-54.1132237453872\\
3.85400000000001	-55.7230314001102\\
4.57599999999999	-57.2145175094209\\
5.37299999999999	-58.6091340050954\\
6.25200000000001	-59.9251765806645\\
7.22	-61.1755411451144\\
8.286	-62.3716957729671\\
9.46100000000001	-63.5235380433838\\
10.757	-64.6386205662245\\
12.188	-65.7234461047343\\
13.771	-66.7841067595181\\
15.523	-67.8243103434264\\
17.464	-68.8476716477432\\
19.616	-69.857006250488\\
22.005	-70.8552246517008\\
24.659	-71.8443064059168\\
27.609	-72.825810728646\\
30.889	-73.8008741674477\\
34.539	-74.7709923813562\\
38.601	-75.7367683072895\\
43.124	-76.6991779410793\\
48.161	-77.6587071116958\\
53.772	-78.6159209929349\\
60.023	-79.5711511544138\\
66.989	-80.5248670890798\\
74.752	-81.4772535289269\\
83.404	-82.4285347859176\\
93.048	-83.3789380480979\\
103.797	-84.3284932230698\\
115.779	-85.2773930773738\\
129.137	-86.2258110543251\\
144.028	-87.1737357944322\\
160.629	-88.1212763068458\\
179.137	-89.0685031302705\\
199.771	-90.0154460653064\\
200	-90.0253971070012\\
};
\addlegendentry{Received power with $f_0$=6 GHz, B=100 MHz}

\addplot [color=mycolor2, line width=1.5pt]
  table[row sep=crcr]{%
0.00100000000000477	22.0158027195581\\
0.0600000000000023	-13.5472222881148\\
0.179000000000002	-23.0412579000398\\
0.355999999999995	-29.0131972398994\\
0.588999999999999	-33.386503176184\\
0.878999999999991	-36.8639747819174\\
1.226	-39.7540066840899\\
1.63	-42.2279493685211\\
2.09299999999999	-44.3995818472157\\
2.61600000000001	-46.3369520734865\\
3.202	-48.0926238321076\\
3.85400000000001	-49.7024314868306\\
4.57599999999999	-51.1939175961413\\
5.37299999999999	-52.5885340918158\\
6.25200000000001	-53.9045766673849\\
7.22	-55.1549412318347\\
8.286	-56.3510958596874\\
9.46100000000001	-57.5029381301042\\
10.757	-58.6180206529449\\
12.188	-59.7028461914547\\
13.771	-60.7635068462385\\
15.523	-61.8037104301468\\
17.464	-62.8270717344636\\
19.616	-63.8364063372084\\
22.005	-64.8346247384212\\
24.659	-65.8237064926372\\
27.609	-66.8052108153663\\
30.889	-67.7802742541681\\
34.539	-68.7503924680766\\
38.601	-69.7161683940099\\
43.124	-70.6785780277997\\
48.161	-71.6381071984162\\
53.772	-72.5953210796553\\
60.023	-73.5505512411341\\
66.989	-74.5042671758002\\
74.752	-75.4566536156473\\
83.404	-76.407934872638\\
93.048	-77.3583381348182\\
103.797	-78.3078933097901\\
115.779	-79.2567931640942\\
129.137	-80.2052111410455\\
144.028	-81.1531358811525\\
160.629	-82.1006763935662\\
179.137	-83.0479032169909\\
199.771	-83.9948461520268\\
200	-84.0047971937215\\
};
\addlegendentry{Received power with $f_0$=3 GHz, B=100 MHz}

\addplot [color=mycolor3, line width=1.5pt]
  table[row sep=crcr]{%
0.00100000000000477	25.5376279006717\\
0.0600000000000023	-10.0253971070012\\
0.179000000000002	-19.5194327189262\\
0.355999999999995	-25.4913720587858\\
0.588999999999999	-29.8646779950703\\
0.878999999999991	-33.3421496008037\\
1.226	-36.2321815029762\\
1.63	-38.7061241874075\\
2.09299999999999	-40.877756666102\\
2.61600000000001	-42.8151268923729\\
3.202	-44.5707986509939\\
3.85400000000001	-46.180606305717\\
4.57599999999999	-47.6720924150277\\
5.37299999999999	-49.0667089107022\\
6.25200000000001	-50.3827514862713\\
7.22	-51.6331160507211\\
8.286	-52.8292706785738\\
9.46100000000001	-53.9811129489905\\
10.757	-55.0961954718312\\
12.188	-56.181021010341\\
13.771	-57.2416816651248\\
15.523	-58.2818852490332\\
17.464	-59.30524655335\\
19.616	-60.3145811560947\\
22.005	-61.3127995573076\\
24.659	-62.3018813115235\\
27.609	-63.2833856342527\\
30.889	-64.2584490730545\\
34.539	-65.228567286963\\
38.601	-66.1943432128963\\
43.124	-67.156752846686\\
48.161	-68.1162820173025\\
53.772	-69.0734958985417\\
60.023	-70.0287260600205\\
66.989	-70.9824419946865\\
74.752	-71.9348284345336\\
83.404	-72.8861096915243\\
93.048	-73.8365129537046\\
103.797	-74.7860681286765\\
115.779	-75.7349679829805\\
129.137	-76.6833859599319\\
144.028	-77.6313107000389\\
160.629	-78.5788512124525\\
179.137	-79.5260780358773\\
199.771	-80.4730209709132\\
200	-80.4829720126079\\
};
\addlegendentry{Received power with $f_0$=2 GHz, B=100 MHz}

\addplot [color=mycolor4, dashdotted, line width=1.5pt]
  table[row sep=crcr]{%
0.00100000000000477	-85.9772291569981\\
200	-85.9772291569981\\
};
\addlegendentry{$\text{Minimum power with B=100 MHz, SNR}_{\text{min}}\text{=0 dB}$}

\addplot [color=mycolor5, dashdotted, line width=1.5pt]
  table[row sep=crcr]{%
0.00100000000000477	-80.9772291569981\\
200	-80.9772291569981\\
};
\addlegendentry{$\text{Minimum power with B=100 MHz, SNR}_{\text{min}}\text{=5 dB}$}

\addplot [color=mycolor6, dashdotted, line width=1.5pt]
  table[row sep=crcr]{%
0.00100000000000477	-75.9772291569981\\
200	-75.9772291569981\\
};
\addlegendentry{$\text{Minimum power with B=100 MHz, SNR}_{\text{min}}\text{=10 dB}$}

\end{axis}
\end{tikzpicture}%

\end{adjustbox}
\caption{Received power varying distance from 0 to 150 km.}
\label{fig:receivedPower}
\end{figure}

\vskip -0.3cm

\definecolor{mycolor1}{rgb}{0.00000,0.44700,0.74100}%
\definecolor{mycolor2}{rgb}{0.85000,0.32500,0.09800}%
\definecolor{mycolor3}{rgb}{0.92900,0.69400,0.12500}%
\begin{figure}[h!]
        \centering
\begin{adjustbox}{width=0.925\columnwidth,keepaspectratio}

\begin{tikzpicture}

\begin{axis}[%
width=\columnwidth,
height=0.8\columnwidth,
scale only axis,
xmin=0,
xmax=20,
xlabel style={font=\color{white!15!black}},
xlabel={Minimum SNR [dB]},
ymin=0,
ymax=400,
ylabel style={font=\color{white!15!black}},
ylabel={Distance [km]},
axis background/.style={fill=white},
xmajorgrids,
ymajorgrids,
legend style={legend cell align=left, align=left, draw=white!15!black, font=\small}
]
\addplot [color=mycolor1, line width=1.5pt]
  table[row sep=crcr]{%
0	125.493605836934\\
0.299999999999997	121.233208226371\\
0.599999999999994	117.117447369841\\
0.900000000000006	113.141412976678\\
1.2	109.30036145627\\
1.5	105.589710258734\\
1.8	102.005032407729\\
2.09999999999999	98.5420512188712\\
2.40000000000001	95.1966351974496\\
2.7	91.964793109369\\
3	88.8426692194222\\
3.3	85.8265386912239\\
3.59999999999999	82.9128031433102\\
3.90000000000001	80.097986356104\\
4.2	77.3787301246286\\
4.5	74.7517902520129\\
4.8	72.2140326790191\\
5.09999999999999	69.7624297449657\\
5.40000000000001	67.3940565755892\\
5.7	65.1060875935372\\
6	62.8957931473243\\
6.3	60.760536254734\\
6.59999999999999	58.6977694567784\\
6.90000000000001	56.7050317784621\\
7.2	54.7799457927287\\
7.5	52.9202147840798\\
7.8	51.1236200084899\\
8.09999999999999	49.388018046343\\
8.40000000000001	47.7113382452344\\
8.7	46.0915802495891\\
9	44.526811614145\\
9.3	43.0151654984586\\
9.60000000000001	41.5548384396782\\
9.90000000000001	40.1440882009308\\
10.2	38.7812316927539\\
10.5	37.4646429650925\\
10.8	36.1927512674671\\
11.1	34.9640391749963\\
11.4	33.7770407780394\\
11.7	32.6303399332998\\
12	31.5225685743006\\
12.3	30.4524050792197\\
12.6	29.4185726941342\\
12.9	28.419838009794\\
13.2	27.4550094901096\\
13.5	26.522936050594\\
13.8	25.6225056850669\\
14.1	24.7526441389805\\
14.4	23.9123136277841\\
14.7	23.1005115988009\\
15	22.316269535136\\
15.3	21.5586518001917\\
15.6	20.8267545214106\\
15.9	20.1197045119137\\
16.2	19.4366582287496\\
16.5	18.7768007665081\\
16.8	18.1393448851018\\
17.1	17.523530070552\\
17.4	16.9286216276611\\
17.7	16.353909803488\\
18	15.7987089405797\\
18.3	15.2623566589512\\
18.6	14.7442130658359\\
18.9	14.2436599922638\\
19.2	13.7601002555586\\
19.5	13.2929569468703\\
19.9	12.6946753305632\\
20	12.5493605836934\\
};
\addlegendentry{Maximum distance with B=100 MHz, $f_0$=6 GHz}

\addplot [color=mycolor2, line width=1.5pt]
  table[row sep=crcr]{%
0	177.474759365695\\
0.300000000000011	171.449647283735\\
0.599999999999994	165.629082460947\\
0.900000000000006	160.006120697674\\
1.19999999999999	154.574053543738\\
1.5	149.326400294946\\
1.80000000000001	144.256900261318\\
2.09999999999999	139.359505297792\\
2.40000000000001	134.628372588517\\
2.69999999999999	130.057857676105\\
3	125.642507727534\\
3.30000000000001	121.377055028668\\
3.59999999999999	117.25641069964\\
3.90000000000001	113.275658623577\\
4.19999999999999	109.430049581457\\
4.5	105.714995586066\\
4.80000000000001	102.126064408323\\
5.09999999999999	98.6589742894308\\
5.40000000000001	95.3095888325379\\
5.69999999999999	92.073912067831\\
6	88.9480836851588\\
6.30000000000001	85.928374428507\\
6.59999999999999	83.0111816468252\\
6.90000000000001	80.1930249958985\\
7.19999999999999	77.4705422861399\\
7.5	74.8404854713428\\
7.80000000000001	72.299716773615\\
8.09999999999999	69.8452049398655\\
8.40000000000001	67.4740216253807\\
8.69999999999999	65.1833379001768\\
9	62.9704208739558\\
9.30000000000001	60.8326304356434\\
9.59999999999999	58.7674161036158\\
9.90000000000001	56.7723139828581\\
10.2	54.8449438254259\\
10.5	52.9830061906995\\
10.8	51.1842797020479\\
11.1	49.4466183966239\\
11.4	47.7679491651323\\
11.7	46.1462692785169\\
12	44.5796439986119\\
12.3	43.0662042699119\\
12.6	41.6041444897033\\
12.9	40.1917203538971\\
13.2	38.8272467759951\\
13.5	37.5090958767044\\
13.8	36.2356950418034\\
14.1	35.0055250459411\\
14.4	33.8171182401313\\
14.7	32.6690568007813\\
15	31.5599710381629\\
15.3	30.4885377623102\\
15.6	29.453478704394\\
15.9	28.4535589916875\\
16.2	27.4875856743083\\
16.5	26.5544063019734\\
16.8	25.652907549074\\
17.1	24.7820138864273\\
17.4	23.9406862981209\\
17.7	23.1279210419191\\
18	22.3427484517529\\
18.3	21.5842317808642\\
18.6	20.8514660842237\\
18.9	20.1435771388906\\
19.2	19.4597204010244\\
19.5	18.7990799983056\\
19.8	18.1608677565629\\
20	17.7474759365695\\
};
\addlegendentry{Maximum distance with B=50 MHz, $f_0$=6 GHz}

\addplot [color=mycolor3, line width=1.5pt]
  table[row sep=crcr]{%
0	396.845626232113\\
0.300000000000011	383.373066044795\\
0.600000000000023	370.357887433595\\
0.899999999999977	357.784562696034\\
1.19999999999999	345.638091281491\\
1.5	333.903981894845\\
1.80000000000001	322.568235207713\\
2.10000000000002	311.617327156605\\
2.39999999999998	301.038192808094\\
2.69999999999999	290.818210771765\\
3	280.945188142308\\
3.30000000000001	271.407345952834\\
3.60000000000002	262.193305122028\\
3.89999999999998	253.292072878379\\
4.19999999999999	244.693029645311\\
4.5	236.385916371533\\
4.80000000000001	228.360822291532\\
5.10000000000002	220.608173101571\\
5.39999999999998	213.118719537109\\
5.69999999999999	205.883526338008\\
6	198.893961588355\\
6.30000000000001	192.141686418196\\
6.60000000000002	185.618645054884\\
6.89999999999998	179.317055212169\\
7.19999999999999	173.229398805581\\
7.5	167.348412983008\\
7.80000000000001	161.667081459785\\
8.10000000000002	156.178626147943\\
8.39999999999998	150.876499069642\\
8.69999999999999	145.754374545134\\
9	140.806141645937\\
9.30000000000001	136.025896904221\\
9.60000000000002	131.407937269701\\
9.89999999999998	126.946753305632\\
10.2	122.63702261581\\
10.5	118.473603494696\\
10.8	114.451528793142\\
11.1	110.565999992343\\
11.4	106.81238147899\\
11.7	103.186195014774\\
12	99.6831143936366\\
12.3	96.2989602804147\\
12.6	93.0296952247\\
12.9	89.8714188439757\\
13.2	86.8203631702845\\
13.5	83.8728881548681\\
13.8	81.0254773254245\\
14.1	78.2747335907957\\
14.4	75.6173751880816\\
14.7	73.0502317673488\\
15	70.5702406092568\\
15.3	68.1744429710951\\
15.6	65.8599805568675\\
15.9	63.6240921072138\\
16.2	61.4641101051029\\
16.5	59.3774575933612\\
16.8	57.3616451002469\\
17.1	55.4142676693953\\
17.4	53.533001990596\\
17.7	51.7156036279789\\
18	49.9599043422977\\
18.3	48.2638095041236\\
18.6	46.6252955948556\\
18.9	45.04240779257\\
19.2	43.51325763983\\
19.5	42.036020790668\\
19.8	40.6089348340588\\
20	39.6845626232113\\
};
\addlegendentry{Maximum distance with B=10 MHz, $f_0$=6 GHz}

\end{axis}

\end{tikzpicture}%

\end{adjustbox}
\caption{Maximum achievable distance varying the SNR threshold.}
\label{fig:maxDistance}
\end{figure}
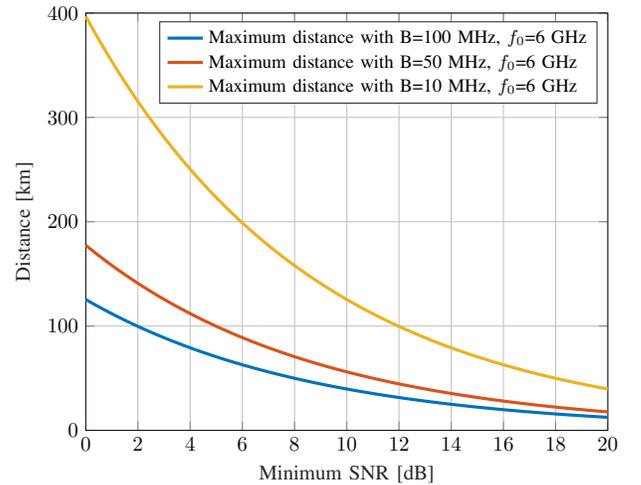

It is important to note that the minimum required SNR is a parameter linked to the \ac{MCS} used for the transmission. The Sidelink standard specifies a list of possible \ac{MCS} indexes. Each \ac{MCS} corresponds to a given target coding rate. Low-index \ac{MCS} ensure greater robustness by introducing redundancy and utilizing low-order modulations, which are more resilient to noise and interference. On the other hand, high-index \ac{MCS} increase spectral efficiency by employing high-order modulations and higher coding rates, but this comes at the cost of reduced robustness and a higher minimum required SNR.

Figures \ref{fig:receivedPower} and \ref{fig:maxDistance} portray the discussed aspects. Fig. \ref{fig:receivedPower} shows the received power for increased considered distance and three received power threshold levels for set \ac{SNR} thresholds, considering 100 MHz bandwidth and three choices for the carrier frequency. From Fig. \ref{fig:receivedPower} it can be seen that distances above 100~km can hardly be supported and can only be achieved considering low \ac{SNR} decoding threshold corresponding to low \ac{MCS} indexes or by reducing the carrier frequency.
Fig. \ref{fig:maxDistance} shows the maximum achievable distance while varying the minimum \ac{SNR} threshold and for different signal bandwidth, for a carrier frequency $f_0$ of 6~GHz. It can be observed that increasing the bandwidth while transmitting the same power increases the contribution of the overall noise. A tradeoff emerges where the optimal bandwidth (and corresponding MCS) maximizes the achievable distance while meeting the SNR requirements for reliable communication.

\subsection{Doppler spread}

In high-speed communication scenarios, such as \ac{A2A} communication, Doppler shift becomes a critical factor to consider.
Doppler shift occurs when there is relative motion between the transmitter and receiver, causing a change in the observed frequency of the signal. The Doppler shift \(\Delta f\) is given by:

\begin{equation}
\Delta f = \frac{v}{c} \cdot f_0
\end{equation}

where \(v\) is the relative velocity between the transmitter and receiver, \(c\) is the speed of light and \(f_0\) is the original frequency of the signal.
It can be observed from the equation, that the frequency shift \(\Delta f\) is directly proportional to the original frequency \(f_0\). Therefore, at higher frequencies, any given relative motion between the transmitter and receiver will result in a larger frequency shift compared to lower frequencies.

Since Doppler shift increases with frequency, selecting the appropriate frequency for communication links is crucial to mitigating Doppler-related issues. However, in addition to Doppler shift, the frequency selection also impacts propagation characteristics, antenna design, interference management, and must comply with spectrum regulations.

Lower frequencies, typically below $3$ GHz, are favored for long-range communication due to their lower free-space path loss and better penetration through obstacles.
Higher frequencies, such as those in the millimeter-wave (mmWave) spectrum, offer larger bandwidths due to more spectrum availability and can, in principle, support higher data rates. However, they require clear \ac{LOS} and they are subject to significant atmospheric absorption and increase the free-space path loss.
Such aspects might be counterbalanced by the possibility for higher frequencies to leverage smaller, more directional antennas, enabling beamforming techniques that enhance signal-to-noise ratio (SNR) and reduce interference in densely populated airspaces. This may be particularly beneficial for applications involving drone swarms or satellite constellations, where precise energy focusing is possible as relative positions are fixed or known.

According to the Sidelink standard specifications
\cite{3GPP_TS_38.101-1}
, NR-V2X can be deployed in specific bands or in concurrency with cellular bands. Nevertheless, the standard admits the possibility for NR-V2X with different bandwidth carrier and up to 100 MHz of bandwidth.
For what regards \ac{A2A} links, all the articles summarized in the survey \cite{channelModSurvey} consider \ac{A2A} links with frequency carrier below 6~GHz. Therefore, in this work, we will consider a carrier frequency allocation around 6~GHz as per 5G standard. 

To manage different levels of Doppler shift in high-speed environments, the NR standard introduced different numerology configurations. Numerologies with higher subcarrier spacing are more robust against the Doppler effect, as Doppler shift is generally evaluated as a percentage of the subcarrier spacing. In the case of OFDM, a Doppler shift level up to 10\% of the subcarrier spacing is often considered tolerable. With 60 kHz of subcarrier spacing, this corresponds to a manageable frequency shift of up to 6 kHz.

For avionic scenarios, the relative speed depends on the type of flight formation and use case. When drones fly together in a tight formation, the relative speed is generally very low (e.g. 10 m/s). These relative speeds do not appear to be a design limitation in terms of Doppler shift.

\definecolor{mycolor1}{rgb}{0.00000,0.44700,0.74100}%
\definecolor{mycolor2}{rgb}{0.46600,0.67400,0.18800}%
\begin{figure}[h!]
        \centering
\begin{adjustbox}{width=0.925\columnwidth,keepaspectratio}

\begin{tikzpicture}

\begin{axis}[%
width=\columnwidth,
height=0.8\columnwidth,
scale only axis,
xmin=0,
xmax=600,
xlabel style={font=\color{white!15!black}},
xlabel={Relative speed [m/s]},
ymin=0,
ymax=15,
ylabel style={font=\color{white!15!black}},
ylabel={Frequency [kHz]},
axis background/.style={fill=white},
xmajorgrids,
ymajorgrids,
legend style={at={(0.03,0.97)}, anchor=north west, legend cell align=left, align=left, draw=white!15!black}]
\addplot [color=mycolor1, dashdotted, line width=1.5pt]
  table[row sep=crcr]{%
0	6\\
800	6\\
};
\addlegendentry{6 kHz Doppler shift threshold}

\addplot [color=black, dashed, line width=1.5pt]
  table[row sep=crcr]{%
300	0\\
300	5.98000000000002\\
};
\addlegendentry{300 m/s}

\addplot [color=mycolor2, line width=1.5pt]
  table[row sep=crcr]{%
0	0\\
800	16\\
};
\addlegendentry{Max Doppler for $f_0$=6.0 GHz}

\end{axis}
\end{tikzpicture}%

\end{adjustbox}
\caption{Doppler shift varying the relative speed.}
\label{fig:Doppler}
\end{figure}
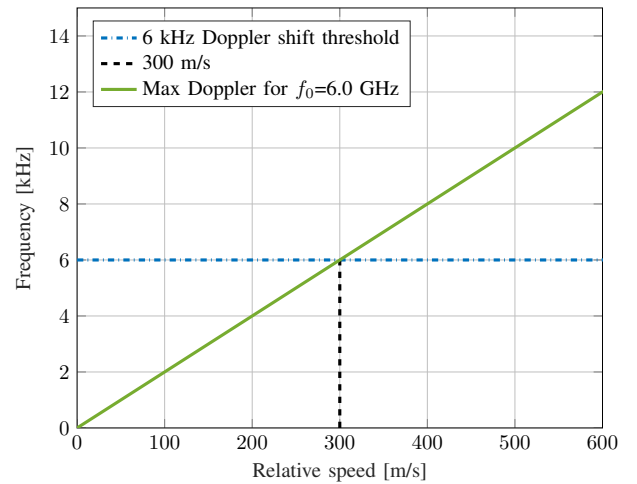

Figure \ref{fig:Doppler} shows the Doppler shift as speed increases. From the figure, it can be seen the Doppler shift is below the threshold for speeds up to 300 m/s, confirming the suitability of 60 kHz subcarrier spacing for high-speed flight dynamics.
However, for higher speeds, the Doppler shift would lead to considerable \ac{ICI}. 
Different approaches may be considered to support speeds higher than 300 m/s. One approach may be to reduce the number of subcarriers that carry data, making the system more robust against ICI at the expense of data rate.
Other solutions may involve considering different modulations, more suitable for high-doppler communication scenarios.

\subsection{Propagation Delay}\label{PropDelay}

The NR-V2X standard has been originally designed for vehicular communications, where network nodes are located at relatively short distances, typically in the order of tens or hundreds of meters. The standard is designed with the assumption of perfect synchronization among all devices. The time axis is rigidly divided into time slots of fixed duration, and transmissions can only last as long as one slot; the start of the slots occurs simultaneously across all nodes, with minimal tolerance for any inaccuracies. This assumption is necessary due to the distributed nature of the system, which enables communication between any two devices.

Given these short distances, the propagation delay in vehicular scenarios is negligible and easily compensated for by the guard interval introduced at the end of each time slot. This guard interval also serves the additional purpose, when necessary, of allowing the radio device to switch from transmission to reception mode. Consequently, a transmission occurring within a given time slot will be received by nearby receivers with minimal delay, ensuring that there is no risk of a transmission being received beyond the end of the time slot.

However, for the scenarios of interest, these assumptions do not hold true. Given the much greater distances considered, on the order of kilometers, it becomes necessary to evaluate the impact of propagation delay.



\begin{figure}[h]
\centering
\includegraphics[width=0.9\columnwidth]{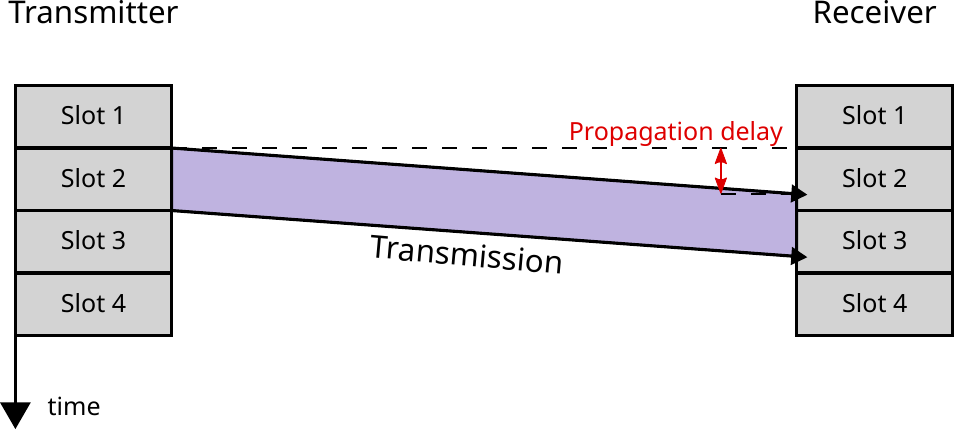}
\caption{Example of propagation delay on slotted transmissions.}
\label{fig:propagationDelay}
\end{figure}

Figure \ref{fig:propagationDelay} illustrates the situation under analysis. The transmitter begins transmission at the start of slot 2; due to the long distance, the receiver, which is synchronized with the transmitter in terms of the start of time slots, receives the transmission with a delay $\tau_p$, equal to the propagation delay. As a result, the transmission is partially received in slot 3. 
A mismatch between the transmission slot and the arrival slot contradicts the assumptions of the standard. This can lead to collisions or errors in sensing and requires a specific evaluation.

A potential countermeasure to the issues caused by propagation delay over long distances can be implemented by leveraging the flexibility introduced in the design of the NR physical layer. Specifically, the duration of the time gap can be configured. By extending the time gap, it becomes possible to tolerate longer delays.

\begin{figure}[h]
\centerline{\includegraphics[width=0.8\columnwidth]{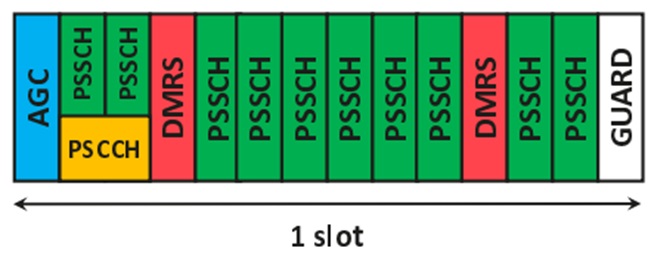}}
\caption{Example of slot format \cite{37.985}.}
\label{fig:slotFormat}
\end{figure}

Figure \ref{fig:slotFormat} illustrates an example of the structure of a slot in NR-V2X. 
A slot comprises 14 OFDM symbols, with the duration of both the slot and each symbol depending on the \ac{SCS}. The figure also shows the physical channels and reference signals, which are multiplexed within a single slot in the time domain and across one or more subchannels in the frequency domain.

The first symbol of any transmission is always reserved for automatic gain control (AGC), which consists of a duplicate of the subsequent symbol. The \ac{DMRS} is used for estimating the radio channel for demodulation. Finally, the \ac{PSSCH} and \ac{PSCCH} carry the payload and the control and scheduling information for sidelink transmissions. The \ac{PSSCH} can be transmitted using anywhere from 5 to 12 symbols, depending on the configuration. Therefore, up to 8 symbols can be left as a guard.

\definecolor{mycolor1}{rgb}{0.49000,0.18000,0.56000}%
\definecolor{mycolor2}{rgb}{0.99000,0.43000,0.22000}%

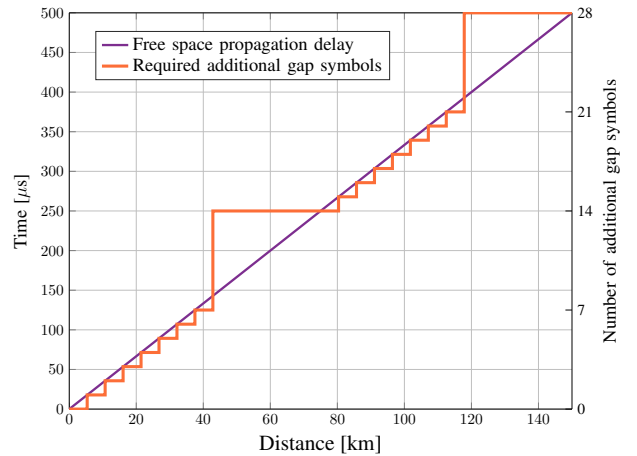
\begin{figure}[h!]
\centering
\begin{adjustbox}{width=0.925\columnwidth,keepaspectratio}

\begin{tikzpicture}

\begin{axis}[%
width=4.521in,
height=3.566in,
at={(0.758in,0.481in)},
scale only axis,
xmin=0,
xmax=150,
xlabel style={font=\color{white!15!black},font=\Large},
xlabel={Distance [km]},
ymin=0,
ymax=500,
ylabel style={font=\large},
ylabel={Time [$\mu$s]},
axis background/.style={fill=white},
axis y line*=left,
xmajorgrids,
ymajorgrids,
legend style={at={(0.03,0.97)}, anchor=north west, legend cell align=left, align=left, draw=white!15!black}
]
\addplot [color=mycolor1, line width=1.5pt]
  table[row sep=crcr]{%
0.000999999999976353	0.0033333333333303\\
150.001	500.003333333333\\
};

\end{axis}

\begin{axis}[%
width=4.521in,
height=3.566in,
at={(0.758in,0.481in)},
scale only axis,
xmin=0,
xmax=150,
ymin=0,
ymax=28,
ylabel style={font=\large},
ylabel={Number of additional gap symbols},
ytick={0,7,14,21,28}, 
yticklabels={0,7,14,21,28}, 
axis y line*=right,
axis x line=none,
every outer y axis line/.append style={black},
every y tick label/.append style={font=\color{black}},
every y tick/.append style={black},
legend style={draw=none}
]
\addplot[const plot, color=mycolor2, line width=2.0pt] table[row sep=crcr] {%
0	0\\
5.35700000000003	1\\
10.714	2\\
16.071	3\\
21.428	4\\
26.785	5\\
32.142	6\\
37.499	7\\
42.856	14\\
80.355	15\\
85.712	16\\
91.069	17\\
96.426	18\\
101.783	19\\
107.14	20\\
112.497	21\\
117.854	28\\
149.996	28\\
};

\end{axis}

\begin{axis}[%
hide axis,
xmin=0,
xmax=1,
ymin=0,
ymax=1,
legend style={at={(0.3,1.4)}, anchor=north west, legend cell align=left, align=left, draw=white!15!black,font=\large}
]
\addlegendimage{line legend, mycolor1, line width=1.5pt}
\addlegendentry{Free space propagation delay}
\addlegendimage{line legend, mycolor2, line width=2.0pt}
\addlegendentry{Required additional gap symbols}
\end{axis}

\end{tikzpicture}%

\end{adjustbox}
\caption{Propagation delay and number of required guard symbols.}
\label{fig:gapSymbols}
\end{figure}

Figure \ref{fig:gapSymbols} illustrates the propagation delay for increasing distance, calculated as the distance divided by the speed of light. As discussed, the propagation delay becomes critical when it exceeds the time gap allocated between transmissions.

For the considered \ac{SCS} the time slot lasts 0.25~ms and consists of 14 OFDM symbols, therefore each symbol has a duration of $17.857~\mu$s, corresponding to 5.357~km at the speed of light. As a result, approximately every $5.3$~km an additional symbol needs to be reserved as a guard. 
The red curve in Figure \ref{fig:gapSymbols} shows the number of additional required guard symbols as the distance increases. 
However, increasing the number of guard symbols reduces the number of symbols available for data transmission, thereby decreasing the system’s data throughput.


For distances beyond $42.4$~km, the required number of guard symbols exceeds the maximum limit of a single slot. As a result, the time gap extends to one or more entire time slots, as represented by the jump in the curve in Figure \ref{fig:gapSymbols}. The standard allows this by providing the flexibility to inhibit transmissions in specific slots using a pre-configured bitmap. In practice, one slot will be used for transmission, followed by one or more unused slots reserved for listening to transmissions from other users. The number of unused slots is determined based on the maximum distance and corresponding propagation delay that the system needs to account for.

\begin{figure}[h]
\centering
\includegraphics[width=\columnwidth]{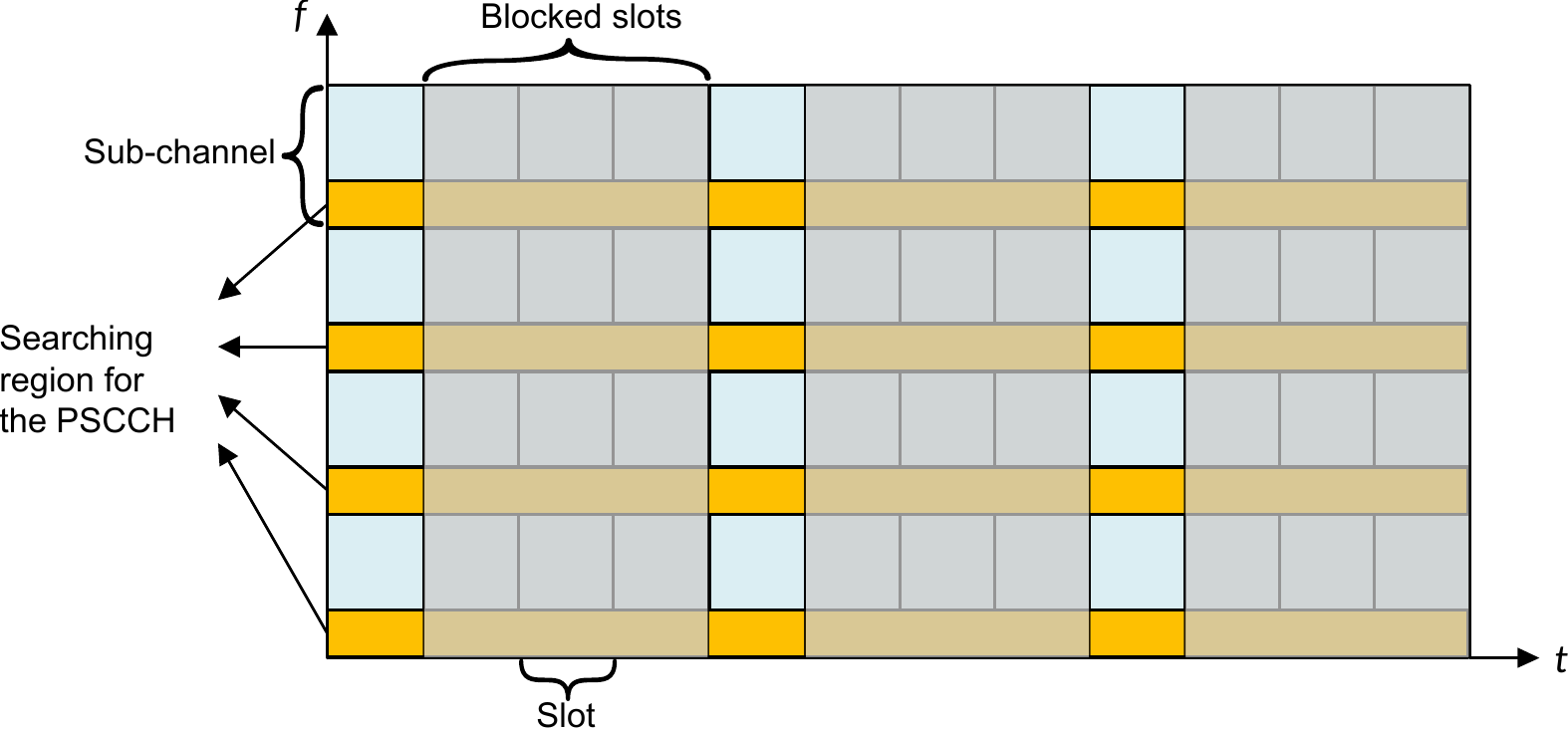}
\caption{Possible slot's subdivision to account for long propagation delay.}
\label{fig:slotBloccati}
\end{figure}

Figure \ref{fig:slotBloccati} portrays an example of such a slot structure. With respect to the solution that inhibits only a few OFDM symbols at the end of the transmission, such a solution poses an important limitation to the amount of data that the communication system can reliably transmit. Essentially the transmitting opportunities are greatly reduced for each slot reserved for the listening phase.


\subsection{Other aspects}

\subsubsection{Synchronization issues}

As the Sidelink has been originally designed for vehicular environments, where GNSS and gNB coverage is available, synchronization is generally assumed to be ideal. However, in the absence of GNSS or gNB, synchronization may still become more complex despite the envisioned multi-layer synchronization procedure.

Moreover, telecommunication systems' dependency on GNSS synchronization has been raising concerns in recent years. Many attacks and vulnerabilities are known and pose risks when autonomous-driven entities are involved. For example, \cite{GNSSattack1} and \cite{GNSSattack2} discuss the possibility for an attacker to control a maritime surface vessel by broadcasting counterfeit civil GNSS signals. Moreover, GNSS systems are always susceptible to jamming.
Therefore airborne Sidelink systems cannot rely solely on GNSS for synchronization purposes.

However, new high-precision clocks recently became available on the market \cite{secondLookCSAC}. \ac{CSAC} are atomic clocks embedded in a chipset. Such devices have similar precision as satellites' atomic clocks and can maintain synchronism for several days before starting to drift apart from the original timing reference \cite{Microchip2020CSAC}. The cost of such devices is in the order of thousands of dollars, making them an attractive option for the scenario of study.

\subsubsection{Other modulations}

The current 4G and 5G communication systems operating at sub-6 GHz frequency bands encounter significant multipath fading. Both LTE and NR standards adopt OFDM as the primary waveform to mitigate this issue. OFDM is also compatible with MIMO, a key technique for achieving high spectral efficiency. However, compared to single-carrier waveforms, OFDM suffers from a higher \ac{PAPR}. 
According to the latest industry forecasts, OFDM is expected to remain relevant in 6G standardization. Nevertheless, alternative waveforms may prove advantageous, particularly in avionic use cases.

One such alternative modulation, which may be well-suited to channels with high Doppler shifts, is \ac{OTFS}. First introduced in \cite{OTFS_Hadani} in 2017, \ac{OTFS} maps each information symbol (e.g., QAM) onto a set of two-dimensional (2D) orthogonal basis functions that span the bandwidth and time duration of the transmission burst or packet. This transforms the time-varying multipath channel into a time-independent 2D channel in the Delay-Doppler domain, ensuring that the communication channel affects all symbols within the same OTFS frame equally \cite{OTFS_Mirabella}.

Another multicarrier waveform worth mentioning is \ac{OCDM}, introduced in \cite{OCDM}. \ac{OCDM} exploits the orthogonality of chirp signals in the Fresnel transform domain, enabling the distinction of overlapping signals in both time and frequency domains. By leveraging chirp signals, which are characterized by a wideband spectrum, \ac{OCDM} can resist various impairments such as phase noise, Doppler effects, and multipath fading \cite{Cuozzo_OCDM}.

Moreover, for \ac{A2A} links, single-carrier waveforms might be advantageous. The adoption of multicarrier systems is typically justified in frequency-selective channels, as they simulate numerous flat sub-channels. However, if the propagation is dominated by a single path, either due to a \ac{LOS} scenario or antenna beam alignment (e.g. at higher frequencies), the likelihood of the channel being frequency-selective is considerably reduced.

\section{Conclusion}\label{sec:conc}

In conclusion, this work has explored the feasibility of extending NR-V2X Sidelink communication to air-to-air links over increasing distances. Various limitations have been identified, regarding communication range, propagation delay, and Doppler spread, with potential solutions and countermeasures proposed. However, the propagation delay at long distances emerges as a critical aspect. Specifically, supporting sidelink communication for distances greater than $42.4~km$ requires to significantly limits the transmission opportunities for users. At such distances, the limitations may outweigh the benefits of synchronized autonomous resource allocation. Despite these challenges, sidelink communication is viable for distances up to $42.4~km$ and does not require modifications to the standard. This result suggests that the Sidelink standard may be more suitable for drone swarm operations in proximity.



\section*{Acknowledgment} This work has been carried out in collaboration with the CNIT National Laboratory WiLab, the University of Bologna, and Leonardo S.p.A.

\bibliographystyle{IEEEtran}
\balance
\bibliography{biblio-japan,biblio-A2A}
\balance

%








\end{document}